\documentclass[%
 reprint,
 amsmath,amssymb,
 aps,
]{revtex4-1}

\usepackage{graphicx}
\usepackage{dcolumn}
\usepackage{bm}
\usepackage{color}

\newcommand{\be}{\begin{equation}}
\newcommand{\ee}{\end{equation}}
\newcommand{\ba}{\begin{align}}
\newcommand{\ea}{\end{align}}
\newcommand{\seff}{s^2_{\mbox{{\tiny eff}}}}
\newcommand{\seffl}{\sin^2 \theta^{\mbox{{\tiny lept}}}_{\mbox{{\tiny eff}}}}
\begin{document}

\title{Comparison of the Standard Theory Predictions of $\bm{M_W}$ and $\seffl$ 
with their Experimental Values}

\author{Andrea Ferroglia}
 \email{aferroglia@citytech.cuny.edu}
\affiliation{%
New York City College of Technology\\
300 Jay Street, Brooklyn, NY 11201 USA
}%

\author{Alberto Sirlin}%
 \email{as6@physics.nyu.edu}
\affiliation{%
 Department of Physics, New York University,\\
 4 Washington Place, New York, NY 10003 USA 
}%

\date{\today}

\begin{abstract}
Assuming that the recently discovered particle at LHC is the Standard Theory (ST) Higgs Boson, we compare the ST predictions of $M_W$ and $\seffl$ with the experimental values of these basic observables. While the $\seffl$ prediction is in excellent agreement with its experimental value, that of $M_W$ shows a $1.33 \sigma$ deviation. Implications of these comparisons for possible future developments at LHC and a future GigaZ linear collider are briefly discussed. It is also pointed out that these comparisons are consistent with the conjecture that the newly discovered particle is indeed the ST Higgs boson.
\end{abstract}

\pacs{Valid PACS appear here}
\maketitle

Assuming that the recently discovered particle at LHC is the Standard Theory (ST) Higgs boson ($H$), the aim of this report is to compare the ST predictions of the basic observables $M_W$ and $\seffl$ with their experimental values. 
These observables are of particular interest because of three reasons: {\em i)} They have been measured accurately, {\em ii)} the theoretical formulas for their calculation include the full one and two-loop electroweak corrections, and  {\em iii)} they play a dominant role in the indirect determination of $M_H$. 
The implications of this comparison  for possible future developments at LHC and a future GigaZ linear collider are briefly discussed.

A crucial  new input is the measurement of the mass of the new particle by the ATLAS collaboration \cite{:2012gk}
\begin{displaymath}
 M_H = 126\pm 0.4 \pm 0.4~\mbox{GeV}\, ,
\end{displaymath}
 and by the CMS collaboration \cite{:2012gu}
\begin{displaymath}
M_H = 125.3 \pm 0.4 \pm 0.5~\mbox{GeV}\, .
\end{displaymath}
Combining the statistical and systematic errors,  we have
\begin{align}
M_H &= 126.0 \pm 0.57~\mbox{GeV} \qquad \mbox{[ATLAS]} \, , \label{atlas} \\
M_H &= 125.3 \pm 0.64~\mbox{GeV} \qquad \mbox{[CMS]} \, . \label{cms}
\end{align}
To the accuracy of our predictions, Eq.~(\ref{atlas}) and Eq.~(\ref{cms}) give the same results, so that either value can be used.

To evaluate $M_W$, we employ the fitting formulas given in Eqs.~(6-8) of Ref.~\cite{Awramik:2003rn}, which include the complete two-loop result and the known higher-order QCD and electroweak corrections. These fitting formulas approximate the full result for $M_W$ to better than $0.5$~MeV for $10~\mbox{GeV} \le M_H \le 1~\mbox{TeV}$. Our other input parameters are:  $M_Z = 91.1876(21)$~GeV \cite{Nakamura:2010zzi}, $M_t = 173.2 \pm 0.9$~GeV \cite{Aaltonen:2012ra}, $\Delta \alpha = \Delta \alpha_h^{(5)} + \Delta \alpha_l = 0.05907 \pm 0.00010$ \cite{Davier:2010nc}, $\alpha_s (M_Z) = 0.1184 \pm 0.0007$ \cite{Beringer:1900zz}. The fitting formulas employ $G_\mu = 1.16637 \times 10^{-5}~\mbox{GeV}^{-2}$ rather than the most recent and accurate value $G_\mu = 1.1663788 \times 10^{-5}~\mbox{GeV}^{-2}$ \cite{Webber:2010zf}. However, this difference is negligible to the accuracy of our calculations.

Inserting the input parameters in the fitting formulas of Ref.~\cite{Awramik:2003rn}, we find the ST prediction:
\begin{align}
M_W^{\mbox{{\tiny (ST)}}} & = 80.361 \pm 0.010~ \mbox{GeV} \, , \label{mwst}
\end{align}
The error in Eq.~(\ref{mwst}) has two components: $\pm 0.006$~GeV arises from the errors in the input parameters, and $\pm 0.004$~GeV is an estimate of the theoretical uncertainty from unknown higher order corrections \cite{Awramik:2003rn}. In Eq.~(\ref{mwst}) we have combined the two errors linearly.

To evaluate $\seff \equiv \seffl$ we employ the fitting formulas given in Eqs.~(48-49) and Table~5 of Ref.~\cite{Awramik:2006uz} which reproduce to high accuracy the complete calculation up to and including two-loop order. Inserting in these formulas the input parameters given above, we find the ST prediction
\begin{align}
s^{2 (\mbox{{\tiny ST}})}_{\mbox{{\tiny eff}}} = 0.23152 \pm 0.00010 \, . \label{seffst}
\end{align}
Again, the error in Eq.~(\ref{seffst}) has two components, combined linearly: $0.00005$ is the parametric error and $0.00005$ is the estimated theoretical error from unknown higher order contributions~\cite{Awramik:2006uz}. 

It is interesting to note that Eqs.~(\ref{mwst}-\ref{seffst}) are very close to the values obtained in a recent global fit of the ST \cite{Baak:2012kk}. This close agreement is related to the observation that $M_W$, $\seff$ and their theoretical expressions play, as mentioned before, a dominant role in the indirect determination of $M_H$.
In fact, the indirect estimates of $M_H$ based on $M_W$ and $\seff$ are close to those obtained from the global fits of all the data (see, for example, Ref.~\cite{Sirlin:2012mh} and refs. cited therein).

Eqs.~(\ref{mwst}-\ref{seffst}) can be compared with the experimental values (see Ref.~\cite{Group:2012gb} and Ref.~\cite{Z-Pole}):
\begin{align}
M_W^{(\mbox{{\tiny exp}})} & = 80.385 \pm 0.015~\mbox{GeV} \, , \label{mwexp} \\
s^{2 (\mbox{{\tiny exp}})}_{\mbox{{\tiny eff}}} &= 0.23153 \pm 0.00016 \, . \label{seffexp}
\end{align}
We remind the reader that, for a long time, there has been an intriguing difference, at the $3 \sigma$ level, between the values of $\seff$ derived from the leptonic asymmetries ($A_{\mbox{{\tiny FB}}}^{0,l}$, ${\mathcal A}_l(P_\tau)$, ${\mathcal A}_l(\mbox{SLD})$) and the hadronic asymmetries ($A_{\mbox{{\tiny FB}}}^{0,b}$, $A_{\mbox{{\tiny FB}}}^{0,c}$, $Q_{\mbox{{\tiny FB}}}^{\mbox{{\tiny had}}}$) \cite{Z-Pole}. In fact, one finds$(\seff)_l = 0.23113(21)$ from the leptonic asymmetries  and $(\seff)_h = 0.23222(27)$ from the hadronic asymmetries.  Since this issue has not been clarified, we follow here the standard procedure of employing the average value derived from all the asymmetries. 

Comparing Eqs.~(\ref{mwst}, \ref{mwexp}) we find a difference
\begin{align}
\delta M_W = M_W^{(\mbox{{\tiny ST}})} - M_W^{(\mbox{{\tiny exp}})} = -0.024 \pm 0.018~\mbox{GeV} \, , \label{dmw}
\end{align}
or equivalently, $\delta M_W = -1.33~\sigma$.

Comparing Eqs.~(\ref{seffst}, \ref{seffexp}):
\begin{align}
\delta \seff = s^{2 (\mbox{{\tiny ST}})}_{\mbox{{\tiny eff}}}  - s^{2 (\mbox{{\tiny exp}})}_{\mbox{{\tiny eff}}}  = -0.00001 \pm 0.00019 \, ,  \label{dseff}
\end{align}
 or, equivalently, $\delta \seff = -0.053 \sigma$.
 
 Thus, we see that the ST prediction of $\seff$ is in excellent agreement with the experimental value, while the ST prediction for $M_W$ shows a $-1.33~\sigma$ deviation.

 For completeness, we note that the ST prediction of Eq.~(\ref{seffst}) differ from $(\seff)_l$ and $(\seff)_h$  (the values obtained separately from the leptonic and hadronic asymmetries) by $s^{2 (\mbox{{\tiny ST}})}_{\mbox{{\tiny eff}}}  - (\seff)_l  = 1.7~\sigma$ and $s^{2 (\mbox{{\tiny ST}})}_{\mbox{{\tiny eff}}} - (\seff)_h   = -2.4~\sigma$.

 In order to decide whether the $\delta M_W$ deviation is a real effect or a statistical fluctuation, it would be very useful to improve the accuracy of the $M_W$ measurement. The question of whether a future measurement of $M_W$ at LHC may be possible with an error $\sim 7~\mbox{MeV}$ \cite{Petersen:2008pn}, or even as low as $5~\mbox{MeV}$ \cite{Peskin}, has been recently discussed. It would also be very useful to decrease the parametric and theoretical errors in Eq.~(\ref{mwst}).  In connection with this remark, we observe that the theoretical error associated with the unknown higher order corrections ($\sim 4$~MeV) is the second largest contribution to the uncertainty in Eq.~(\ref{mwst}), after the parametric one arising from the experimental error in  the top quark  mass ($\sim 5$~MeV).
 
 A future GigaZ linear collider may be able to measure $\seff$ with an error of $\sim 1 \times 10^{-5}$. In order to find out whether the current agreement between the ST prediction of $\seff$ and its experimental value will survive, or a deviation emerge, a decrease of the parametric end theoretical errors in Eq.~(\ref{seffst}) by a factor of $\approx 10$ will be required. We observe that, in the calculation of $\seff$, the theoretical error associated with the unknown higher order corrections is already the largest contribution to the uncertainty in Eq.~(\ref{seffst}).
 
 It is important to note that the excellent agreement of the $\seff$ comparison and the fact that the $\delta M_W$ deviation is relatively small, of ${\mathcal O} (1 \sigma)$, are consistent with the conjecture that the newly discovered particle is indeed the Higgs boson, the fundamental missing piece of the ST.
 
The authors are indebted to Kyle Cranmer for very useful information. 
The work of A.~Ferroglia was supported in part by the PSC-CUNY Award N. 64133-00 42 and by the National Science Foundation Grant No. PHY-1068317.
The work of A. Sirlin was supported in part by the National Science Foundation Grant No. PHY-0758032.

\bibliography{MwFixedMh}

\end{document}